\documentstyle[epsf,rotate]{mn}
\title{B$-$R Colours of Globular Clusters in NGC 6166 (A2199)}

\author[T.J. Bridges et al.]
{T.J. Bridges,$^{1,4}$, D. Carter$^1$\thanks{Visiting Astronomer,
William Herschel Telescope.
The WHT is operated on the island of La Palma by the Royal Greenwich
Observatory at the Spanish Observatorio del Roque de los Muchachos of
the Instituto de Astrofisica de Canarias.}
W.E. Harris,$^2$ C.J. Pritchet,$^3$ \\ 
$^1$Royal Greenwich Observatory,
Madingley Road,
Cambridge, England,
CB3 0EZ\\
$^2$Dept. of Physics \& Astronomy, McMaster University
Hamilton, Ontario, Canada\\
$^3$Dept. of Physics, University of Victoria
Victoria, B.C., Canada \\
$^4$E-mail:~~tjb@ast.cam.ac.uk}

\date{Received:~ Nov. 17, 1995 ~~~Accepted:~ Feb. 5, 1996}

\pubyear{1995}

\begin{document}

\maketitle

\begin{abstract}

We have analysed new R-band photometry of globular clusters in
NGC 6166, the cD galaxy in the cooling flow cluster A2199.  In
combination with the earlier B photometry of Pritchet \& Harris
(1990), we obtain
B$-$R colours for $\sim$ 40 globular clusters in NGC 6166.  The
mean B$-$R is 1.26 $\pm$ 0.11, corresponding to a mean 
[Fe/H] = $-$1 $\pm$ 0.4.  Given that NGC 6166 is one of the most
luminous cD galaxies studied to date, our result implies significant 
scatter in the relationship between mean cluster [Fe/H] and 
parent galaxy luminosity.  We obtain a globular cluster specific
frequency of S$_N$ $\sim$ 9, 
with a possible range between 5 and 18.  This value is inconsistent
with the value of S$_N$ $\leq$ 4 determined earlier by 
Pritchet \& Harris (1990) from B-band photometry, 
and we discuss possible reasons for
the discrepancy.   Finally, we 
reassess whether or not cooling flows are
an important mechanism for 
forming globular clusters in gE/cD galaxies. 

\end{abstract}


\section{Introduction}

Broadband colours provide a relatively easy way to estimate 
metallicities of extragalactic globular clusters. 
The mean cluster metallicity (we will use colour and
metallicity interchangeably in this paper, implicitly assuming that 
we are dealing with old stellar populations where age effects are
not important) gives the overall level of chemical enrichment in the
globular cluster system
(GCS).  There appears to be a relationship between mean cluster 
[Fe/H] and parent galaxy luminosity (e.g. Brodie 1993) in the sense
that more luminous galaxies have, on average, more metal-rich globular
clusters.  However, recent work (e.g. Zepf, Ashman, \& Geisler
1995a; Secker et
al 1995) has shown that there is considerable scatter about  
this relationship. 

The cluster metallicity {\it distribution} (MD) provides far more
information than the mean metallicity alone.
The width of the MD is presumably indicative of the
inhomogeneity of the protogalactic gas from which clusters formed and/or their
subsequent chemical enrichment.  In addition, multimodality in cluster 
MDs has been detected in several 
ellipticals (e.g. M87: Lee \& Geisler 1993,
Whitmore et al 1995; NGC 3311: Secker et al 1995;
NGC 1399: Ostrov, Geisler, \& Forte 1993; NGC 3923: Zepf, Ashman, \& Geisler
1995a),
and is most naturally explained by
distinct epochs of cluster formation.  The existence of
spatial metallicity gradients is also interesting, since 
such gradients are predicted by many cluster formation scenarios,
including classical dissipative collapse (e.g. Eggen, Lynden-Bell,
\& Sandage 1962), galaxy
mergers (Ashman \& Zepf 1992; Zepf \& Ashman 1993), and possibly 
cooling flows
if clusters have been forming out of the (subsonic) flow over
several Gyr (e.g. Fabian 1994).

The number of early-type galaxies with measured globular cluster colours
is still small ($<$ 20).  NGC 6166, the central cD in
Abell 2199, is an interesting addition, given that it is one of
the most luminous galaxies known (M$_V$ $\simeq$ $-$23.6 for
H$_0$=75).  It is also very X-ray luminous, together
with a substantial cooling flow of 100--150 M$_{\odot}$/yr
(Edge, Stewart, \& Fabian 1992).
NGC 6166 has long been regarded as the classic multiple-nucleus
galaxy, though two of the `nuclei' are in fact not bound to
the brightest one (Tonry 1984; Lachieze-Rey, Vigroux, \& Souviron 
1985; Lauer 1986).  
Lachieze-Rey et al.
and Peletier (private communication)
both find a colour gradient in NGC 6166, with
$\delta$(B$-$R) $\simeq$ $-$0.4 from the centre
out to $\sim$ 1 arcmin from the
galaxy centre (see Figure 5).
Cardiel, Gorgas, \& Aragon-Salamanca (1995) also find stellar
metallicity gradients from long-slit spectroscopy, and there is
some evidence  from Einstein and Ginga data of a metallicity
gradient in the X-ray gas.  

Two of us (Pritchet \& Harris 1990; hereafter PH) 
previously detected a GCS in
NGC 6166, using B-band images taken at the CFHT.  The determination
of the globular cluster specific frequency S$_N$ was hampered by the
lack of a background frame, but PH estimated that S$_N$ $\leq$ 4.
In this respect then, NGC 6166 is clearly unlike M87 and some other central
cD/gE 
galaxies where S$_N$ $\simeq$ 15--20 (e.g. Harris 1991).  PH used this
result, together with the large cooling flow, to argue that most
globular clusters in central cD/gE galaxies do not form from
cooling flows.  We will return to this point in our Discussion
(Section 4).
The motivation for the present R-band study is two-fold:
first, to confirm or not the specific frequency found by PH, and
second to obtain B$-$R colours and thus metallicities
for the brightest globular
clusters in NGC 6166.

\section{Data and Analysis}

The data for this study were obtained by D. Carter in August 1988 during
the commissioning of the 4.2m William Herschel Telescope (WHT) in La Palma.
The observations consist of a series of 12~$\times$~300 sec dithered 
exposures for a field centred on NGC 6166 (Figure 1), and a field offset
$\sim$ 10 arcmin from the galaxy (the `background' field).  Data were taken
in the Kron-Cousins R-band using a GEC CCD (0.27 \arcsec/pixel, R/N $\sim$ 
9 e$^-$, 1.7 $\times$ 2.6 arcmin). 
After the usual
pre-processing (bias subtraction, flat-fielding), the galaxy and
background frames were median-combined into a single effective 
exposure of 3600 sec each; the FWHM on the combined image is 
0.9\arcsec~  for the galaxy and background fields, and
the effective field size is 1.5 $\times$ 2.4 arcmin 
for each field.  The STSDAS ISOPHOTE and BMODEL
routines were then used to fit and subtract most of the smooth halo
light of NGC 6166.  The area within 15$\arcsec$ of the galaxy centre,
and smaller areas around other galaxies and bright stars in the
frame, were masked out and not used in the following analysis; the
central masked region includes the superposed `multiple nuclei'.

Detection and photometry of objects in the combined galaxy and
background fields were done with the IRAF version of DAOPHOT.
Image moments were also calculated 
to allow rejection of non-stellar images in a uniform way
(e.g. Butterworth \& Harris 1992).   46 and 181 objects with
reliable classifications were rejected as non-stellar in the
background and galaxy frames, respectively, to R $\sim$ 26--26.4
(the precise magnitude limit is difficult to determine, since our
photometric calibration applies only to {\it stellar} objects).
From the deep R-band galaxy counts of Smail et al. (1995), we
expect $\sim$ 250 galaxies in each of the galaxy and program fields
to R = 26, within our $\sim$ 3 arcmin$^2$ field (which includes a
correction for the masked-out regions).  The agreement with the
numbers of rejected objects in each field is as good as can be
expected, given the uncertainties as well as possible small-scale
clustering of background galaxies.  A further check of our image
classification 
comes via a comparison with PH.  There are only 13 objects
(out of 372 total) from our final `stellar' R list that were
classified by PH as extended objects from their B photometry. 
Visual examination shows that  
about half of these 13 objects do appear to be
stellar in our R frame.  In any case, these results
show that the image classification has been performed reliably in
both the B and R datasets.

Variable extinction due to dust causes
an uncertainty in the zero point as determined from the R band
standards.  However, we have refined the zero point using surface
photometry of NGC 6166 and NGC 6173 (for the background 
field) by R. Peletier (private communication).  The
uncertainty in this zero point is of order 0.1 mag. 
Finally,
determination of photometric incompleteness and errors was done 
by using ADDSTAR to put in groups of 100
PSF stars at several magnitudes, and rerunning DAOPHOT exactly
as for the original frame.  The 50\% completeness level occurs
at R $\sim$ 24.85 (25.15) for the galaxy (background)
field; the magnitude {\it scale}
error is $\simeq$ 0.14 (0.05)  mag at this level, while the {\it rms}
uncertainty reaches $\simeq$ 0.30 (0.20) mag.  Before correcting for
incompleteness, we find 304 objects classified as stellar
to R=25.0 in the galaxy field 
(at R=25.0 the completeness is $\sim$ 35\% in the galaxy
field).  We have looked for possible variations of the photometric
completeness with distance from the galaxy center, by adding in
1000 stars and determining the completeness in 25-pixel wide
annuli with mean radii of 100, 150, and 200 pixels ($\sim$ 0.5,
0.75 and 1 arcmin); this experiment was run for three different
magnitudes corresponding to completeness levels between 0.3 to 0.9.
There is no trend for a variation of the completeness with radius,
and no statistically significant differences in the completeness
levels at the 3 radii, at any magnitude.

\begin{figure}
\epsfysize 4.5truein
%\hfil\epsffile{fig1.ps}\hfil
%\vspace{302pt}
\caption{R image of NGC 6166, with
North to the top and East
to the left.  The galaxy has been fit and subtracted
using the STSDAS ISOPHOTE package within IRAF.  The center of NGC 6166
has been marked with an `X'; the object just North-West of the galaxy
center is a superposed cluster galaxy.}
\end{figure}

\section{Results}

\subsection{Radial Profile of Globular Cluster System}

The radial profile of the NGC 6166 GCS is presented in Figure 2,
with Log (number density) vs Log (radius in arcmin); the solid line is
the best-fit least-squares line.  To produce this Figure, we have put
the 304 stellar objects to R=25.00 into bins of width 25.0
pixels (6.75\arcsec), and corrected each bin for photometric incompleteness.
Finally, we have subtracted the density of background objects to R=25.00
(26 $\pm$ 3 per arcmin$^2$), as determined from our background field.
The NGC 6166 GCS appears to be one of the shallowest found to date, with
a best-fit slope of $-$0.95 $\pm$ 0.1, though this is based on a small
number of clusters.
This result is consistent with the finding by Harris (1993) of a 
correlation between GCS profile slope and parent galaxy luminosity,
in the sense that the GCS exhibits a flatter profile in more luminous
galaxies.

\begin{figure}
\epsfysize=3.5truein
\hspace{-50pt}\epsffile{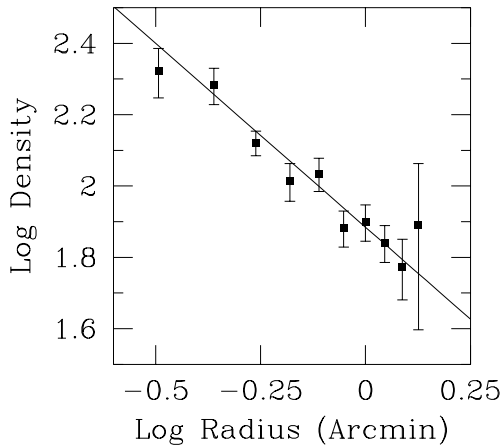}
%\vspace{302pt}
\caption{The Radial Density Profile of the NGC 6166 GCS.
Log Density
is plotted against Log Radius (Arcmin).  The solid line is the
least-squares fit, with a slope of $-$0.95.}
\end{figure}

%\placefigure{fig2}

\subsection{Globular Cluster Luminosity Function}

The NGC 6166 globular cluster luminosity function (GCLF) is shown 
in Figure 3 (filled squares),
as Log(Number) vs. apparent R mag.  The data have again been
background-subtracted, and corrected for photometric incompleteness;
the completeness is shown above the x axis.
The open circles in Figure 3 represent the M87 GCLF, taken from
McLaughlin, Harris \& Hanes (1994; error bars have been omitted
for clarity).  The M87 data have been converted from V to R taking
V$-$R~=~0.55, and shifted in distance modulus by $\Delta$(m$-$M)=4.3
mag (see Section 3.3); no vertical scaling has been applied to either
set of data.  It can be seen that the NGC 6166 GCLF is quite 
consistent with the (shifted) M87 GCLF.  Unfortunately, the NGC 6166
GCLF spans too small a range to even contemplate solving for the
turnover magnitude by Gaussian fitting; to R=25.20 we are only detecting
$\sim$ 5\% of the NGC 6166 GCS.

\begin{figure}
\epsfysize 3.5truein
\hspace{-50pt}\epsffile{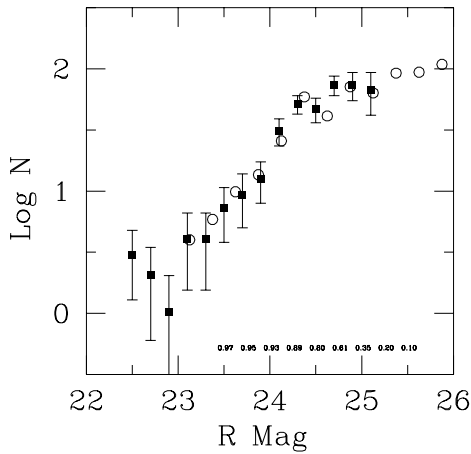}
%\vspace{302pt}
\caption{The Luminosity Function of the NGC 6166 GCS
(filled squares).  Log Number
is plotted against R mag, with photometric completeness shown above the
x axis.  Open circles are the M87 GCLF, taken from McLaughlin,
Harris, \& Hanes 1994, converted from V to R assuming V$-$R=0.55,
and shifted by $\Delta$(m$-$M)=4.3 mag (error bars are omitted for
clarity).  No vertical scaling has been applied to either dataset.}
\end{figure}

%\placefigure{fig3}

\subsection{Globular Cluster Specific Frequency}

We have computed a {\it local} globular cluster specific frequency S$_N$
for NGC 6166, i.e. {\it within} our CCD field (corresponding to a
radius of $\sim$ 40 kpc for H$_0$=75), excluding masked-out regions
around the centre of NGC 6166 and other bright objects in the field.
The total number of clusters over all magnitudes is determined by  
assuming that the NGC 6166 GCLF is intrinsically the same 
as that of the Virgo gEs (e.g. same absolute peak magnitude and
dispersion). 
From Lucey et al (1991), $\Delta$(m$-$M) = 0.7 $\pm$ 0.1 between
Coma and A2199, and from Bower, Lucey, \& Ellis (1992), the offset between
Coma and Virgo is 3.69 mag, 
giving $\Delta$(m$-$M)$_{A2199-Virgo}$=4.4.
On the other hand, the redshift ratio gives
a difference of 4.2, and we adopt  the average of these two estimates as
$\Delta$(m$-$M)$_{A2199-Virgo}$=4.3 $\pm$ 0.1. 
Within our CCD field, there are 385 $\pm$ 40 stellar objects to R=25.2
after correction for photometric incompleteness and background
subtraction,
or to B=26.45 $\pm$ 0.1 assuming B$-$R=1.25 (see next section).
This limiting magnitude corresponds then to B=22.15 $\pm$ 0.15 at
Virgo.  Given that the GCLF of the Virgo gEs peaks at B=24.7 $\pm$ 0.3
(Harris et al 1991), we are 2.55 $\pm$ 0.34 mag short of the peak
in the NGC 6166 GCLF.  Assuming that the dispersion is 1.4 $\pm$ 0.1,
again as found for the Virgo gEs (Harris et al 1991), we find the
total number of clusters over all magnitudes to be 11,000 with a
possible range between 6200 and 22,000.  The large 
uncertainty in 
N$_{tot}$ is almost entirely due to the uncertainties in the 
peak magnitude and dispersion in the adopted Virgo GCLF, amplified 
by the fact that we sample such a small fraction ($\sim$ 3.5\%) of
the NGC 6166 GCLF.
  
The total galaxy magnitude in the CCD field is determined by simply
adding up all of the counts within the frame
(with stellar objects subtracted, and again
excluding the masked-out regions), subtracting the background
sky, and using our photometric calibration to convert to true 
R. 
This yields R=12.27 $\pm$ 0.1.  From RC3, (B$-$V)$_{0,T}$=0.91 $\pm$ 
0.05, and from Peletier's unpublished photometry, (B$-$R)=1.35 $\pm$
0.05.  Thus (V$-$R)=0.44 $\pm$ 0.07, and V=12.71 $\pm$ 0.12.  Next 
we find the A2199 distance using
$\Delta$(m$-$M)$_{A2199-Virgo}$=4.3 (above), and 
the Virgo distance of 17 $\pm$ 2 Mpc from the recent
Cepheid result of Freedman et al (1994), which gives then the
A2199 distance as 35.45$^{+0.26}_{-0.29}$.  Thus,
M$_{V,T}$=$-$22.74$^{+0.29}_{-0.31}$. 

Finally, the globular cluster specific frequency is given by:

\begin{equation}
S_N=N_T10^{0.4(M_{V,T}~+~15)},
\end{equation}

\noindent or S$_N$=9 with a possible range between 5 and 18, where the 
uncertainty in S$_N$ is totally dominated by the
uncertainty in N$_T$.  

Our result is {\it not} consistent with that
found by PH, namely S$_{N,local}$ $\leq$ 4 (note that PH also used
B$_0$=24.7 and $\sigma$=1.4 for the Virgo GCLF).
There are two possible  reasons for this
discrepancy.
First, PH did not have a separate background field, and their
background may have been contaminated by the outer part of the GCS.
Second, PH found evidence for a population of extended objects,
perhaps dwarf galaxies, in their B data.  Although we have not found
any evidence for such a population in our R data, the effect would
be to lower our S$_N$ value, since in that case our background
would be underestimated; at the same time, PH would have
overestimated their background and hence underestimated their S$_N$.
Some significant fraction of our residual objects might be 
unresolved dEs, since
we cannot resolve objects smaller than $\sim$ 0.6 kpc in size at the
distance of NGC 6166 with our $\sim$ 1$^{\prime \prime}$ seeing.
For instance, in the Coma cluster, Bernstein et al (1995) find that
the faintest dEs in their sample for which they can
reliably determine sizes (between 22.5 $<$ R $<$ 23.0, which
would translate $\sim$ 0.5 mag fainter in A2199) have half-light
radii between 0.4--0.75 kpc; r$_{1/2}$ decreases with 
magnitude and objects 0.5 mag fainter than this are roughly a factor of 
two smaller. 

However, this {\it local} value
of S$_N$ may well differ from the {\it global} value.  The main
uncertainty, as discussed in PH, is the unknown behavior of the
globular cluster spatial distribution outside the CCD field.  If
the clusters are more spatially extended than the halo light (which
is certainly true {\it inside} the CCD field) at all radii, then S
will {\it increase} with radius.  PH estimate that
S$_{global}$/S$_{local}$ = 1.6 if NGC 6166 is similar to M87
in this respect.  In this context, it is interesting to compare
our local value of S$_N$ for NGC 6166 with 
a similar one for M87.
We have used the globular cluster counts of McLaughlin, Harris, 
\& Hanes 
(1994) and the photometry of Carter \& Dixon (1978) and Boroson,
Thompson, \& Schectman
(1983) to compute S$_N$ for M87 out to 7 arcmin radius;
this corresponds to 
$\sim$ 35 kpc for a Virgo distance of 17 Mpc, roughly the same
physical radius as our data for NGC 6166.  With the same GCLF
parameters and Virgo distance as assumed above, we find that 
S$_{N,M87}$=11.7 $\pm$ 1.7 between 1.2 to 7 arcmin 
radius (McLaughlin et al did not carry out cluster counts
within 1.2 arcmin of the galaxy centre).  This 
value is not so different from what we have found for NGC 6166
above.  Thus, the present data do not allow us to rule out the 
possibility of NGC 6166 being
a high--specific frequency system.
We plan to obtain 
deeper, wider-field data so that a better global S$_N$ value
can be determined.

\subsection{B$-$R Colours of NGC 6166 Globular Clusters}

B$-$R colours were determined by matching objects from the earlier 
B data of PH with the present R data; only objects classified as stellar
in {\it both} lists were used.  The transformation between the two
coordinate systems (including a small rotation of $\sim$ 4 deg) was
obtained from several bright stars in common.

In Figure 4, we show the R, B$-$R CMD for 
those (70) objects with 24.0 $\leq$ R $\leq$ 25.0
and 0.0 $\leq$ B$-$R $\leq$ 2.0; in the lower right hand corner is
a representative error bar.  
To isolate a cleaner sample of
NGC 6166 globular clusters, we applied magnitude
and colour limits.  The magnitude limits used were 
24.0 $\leq$ R $\leq$ 24.5, where the bright end is set at the expected
onset of the GCS, and the faint end is set at a completeness of
60\%.  The colour limits are 0.8 $\leq$ B$-$R $\leq$ 1.6, which
generously includes all known Galactic globular clusters (Harris 1995). 
The dashed and solid lines in Figure 4 
show our cuts in B$-$R and R respectively.
With these limits (ie. within the central box), there are 37
objects.
In Figure 5, 
we plot B$-$R 
against galactocentric radius for these 37 objects (open circles); the
solid horizontal line shows the mean colour of these objects
(see below), and the filled squares represent the colour profile
of the galactic halo light from Peletier (priv. comm.).  While
there is a colour gradient in the halo light, none is apparent in
the globular clusters.  However, a small gradient could easily be
masked by the small number of clusters and our large uncertainties.

\begin{figure}
\epsfysize 3.5truein
\hspace{-50pt}\epsffile{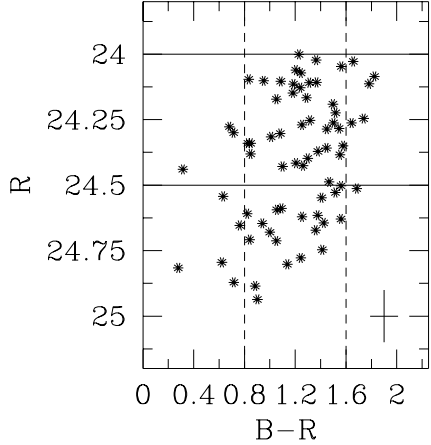}
%\vspace{302pt}
\caption{The R,B$-$R CMD of the NGC 6166 GCS.  The region
enclosed
by the dashed and solid lines is where the globular clusters
are expected to lie; there are 37 objects in this
region. At the lower right hand corner is a
representative error bar.}
\end{figure}

\begin{figure}
\epsfysize 3.5truein
\hspace{-50pt}\epsffile{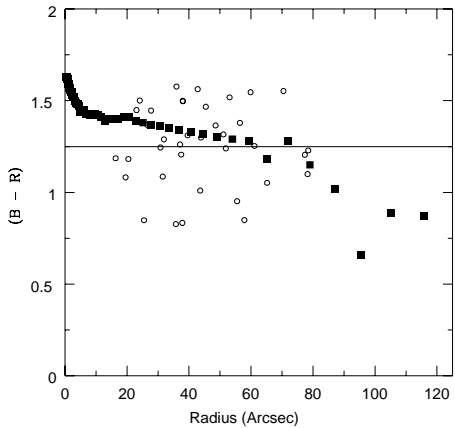}
%\vspace{302pt}
\caption{B-R vs. Galactocentric Radius.  B-R (open circles)
is plotted against
Galactocentric Radius (arcsec); the solid line shows the mean colour
of these objects. The filled squares are the colour profile
of the halo light from Peletier.}
\end{figure}

%\placefigure{fig4}

%\placefigure{fig5}

The B$-$R histogram of the NGC 6166 `clusters' is shown in Figure 6.
In this Figure, we show those (44) objects with 
24.0 $\leq$ R $\leq$ 24.5, and 0.0 $\leq$ B$-$R $\leq$ 2.0.
Again, these data are consistent with a globular cluster population,
with a roughly Gaussian distribution and a mean B$-$R $\sim$ 1.25.
Within the small numbers there is no evidence for multimodality,
as has been found for clusters in other ellipticals (e.g.
Zepf, Ashman, \& Geisler 1995a; Secker et al 1995).  However, Ashman,
Bird, \& Zepf (1994)
showed that it is essentially impossible to detect bimodality in
datasets with N $<$ 50.

\begin{figure}
\epsfysize 3.5truein
\hspace{-50pt}\epsffile{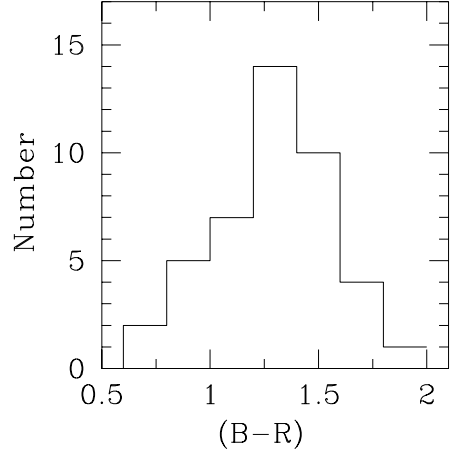}
%\vspace{302pt}
\caption{B-R Histogram of NGC 6166 Globular Clusters.
There are
44 objects with 24.0 $\leq$ R $\leq$ 24.5 and 0.0 $\leq$ B$-$R $\leq$ 2.0.}
\end{figure}

%\placefigure{fig6}

In order to quantify the mean B$-$R colour, we have calculated the
mean and median B$-$R for those (37) objects with
24.0 $\leq$ 24.5 and 0.8 $\leq$ B$-$R $\leq$ 1.6, 
finding both the mean and median B$-$R to be 1.26 $\pm$ 0.11  
($\pm$ 0.10 of this comes from the possible systematic error in our
R zeropoint [Section 2], while the uncertainty in the B zeropoint is
$\pm$ 0.05 [PH]).  The foreground extinction towards NGC 6166 is
negligible, and we adopt E(B$-$V)=0.00 $\pm$ 0.015 mag
(Burstein \& Heiles 1984).
Using the conversion between (B$-$R)$_0$ and
[Fe/H] established for Galactic globular 
clusters by Reed, Harris, \& Harris (1994), this 
corresponds to [Fe/H]$_{mean}$=$-$1.05 $\pm$ 0.40.  The uncertainty in
[Fe/H]
includes both the photometric error in our mean (B$-$R)$_0$ 
($\pm$ 0.11 mag), and the scatter in Reed et al.'s 
relation between (B$-$R)$_0$ and [Fe/H] ($\pm$ 0.21 dex in [Fe/H];
their equation 3c).
Our incompleteness
does not have a steep colour dependence, so the mean and median
B$-$R colours are unchanged by the application of techniques like 
that of Zepf, Ashman, \& Geisler (1995a) to account for colour-dependent
incompleteness.

We have also experimented with relaxing the colour and magnitude limits
given above between 24 $\leq$ R $\leq$ 25 and 0.0 $\leq$ B$-$R $\leq$
2.0.  While the number of matched objects increases to 70 in the
most `relaxed' case, the corrected median B$-$R only changes by $\pm$ 0.05.
We summarize by stating that [Fe/H]=$-$1 $\pm$ 0.4.
This mean cluster [Fe/H] is inconsistent with the metallicity of the
A2199 ICM, which
was recently determined by White et al (1994) to lie between
[Fe/H]=$-$0.2 and $-$0.5, 
based on Ginga and Einstein X-ray data; see further discussion of
this point in Section 4.

It is also interesting to compare the NGC 6166 mean cluster metallicity with
that of clusters in other galaxies.  In Figure 7, we show  
mean cluster [Fe/H] plotted against
parent galaxy luminosity for 17 galaxies of all
types (this Figure is adapted from Secker et al 1995); 
our value for NGC 6166 is shown as a triangle.
Recent data (e.g. NGC 3923: Zepf, Ashman, \& Geisler 1995a; NGC 3311: Secker 
et al 1995; NGC 6166: present work) have tended to {\it increase}
the scatter about any possible trend at the high-luminosity end.
Such variation probably indicates that cluster formation in ellipticals
is more complicated than predicted by monolithic collapse models, and
other effects, including mergers or multiple formation and
enrichment stages, are also important; see Ashman \& Bird (1993) 
and Perelmuter (1995) for further discussion.

\begin{figure}
\epsfysize 3.5truein
\hspace{-50pt}\epsffile{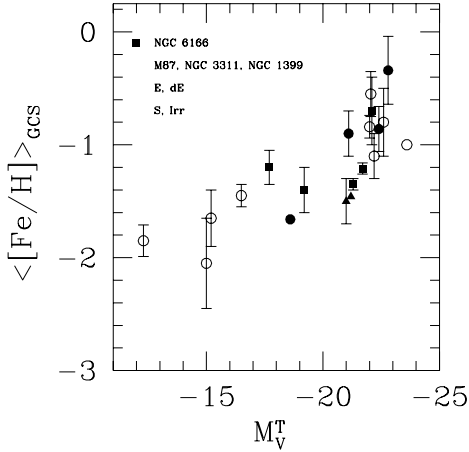}
%\vspace{302pt}
\caption{Mean [Fe/H] metallicity of globular
cluster systems as a
function of parent galaxy luminosity.  This Figure has been adapted
from Figure 7 of Secker et al (1995).  The data for the S, Irr, E, and
dE galaxies are taken from Harris (1991).  Globular cluster data for
NGC 1399 from Ostrov et al (1993), for M87 from Lee \& Geisler (1993),
and for NGC 3923 from Zepf et al (1995a).  Globular cluster [Fe/H] for
NGC 6166 from the present work, and galaxy photometry from RC3 with
H$_0$=75 assumed.}
\end{figure}

%\placefigure{fig7}

\section{Discussion: ~The Role of Cooling Flows}

What constraints can these and similar data place on the formation
of globular clusters in cooling flows?  Here we discuss several 
arguments that have been raised 
against forming clusters in cooling flows. 

\noindent {\bf 1:}  {\it There is no correlation between cluster
specific frequency S$_N$ or total number of clusters,
and properties of the X-ray gas (X-ray luminosity, gas temperature,
total gas mass, or cooling flow rate).}  

\noindent Some cDs (e.g. M87, NGC 3311, NGC 4874) have
huge cluster populations and low cooling flow rates, while others
such as NGC 4073 (MKW 4) 
have `normal' cluster systems and similarly low cooling flow rates.
If cooling flows are putting the majority of their mass into
low-mass objects (LMOs), and a small fraction into globular
clusters, then S$_N$ should *increase* with 
mass-flow rate since the LMOs basically have no effect on the
galaxy luminosity, whereas more massive cooling flows would
form a larger number of (luminous and hence observable) clusters. 
The lack of a correlation
between N$_{tot}$ and ICM properties is even harder to reconcile with
a cooling flow scenario.  While it should not be forgotten that N$_{tot}$
and S$_N$ (and the X-ray and cooling flow parameters) could be uncertain
by factors of two or more, these numbers already cover such a huge
intrinsic range that it is very hard to believe that any well-defined
correlation could have been obscured just by observational scatter.
Deeper imaging, and hence a better-constrained
S$_N$ value for NGC 6166, will be very useful, 
given its large cooling
flow rate of $\sim$ 150 M$_{\odot}$ per yr.
See Harris, Pritchet \& McClure (1995) for a very thorough discussion
of these points.

\noindent{\bf 2:~~}
{\it If cooling flows have been
forming globular clusters continuously to the present day, we would expect
to see some bright, blue young clusters formed recently.}

\noindent While young clusters appear to have been found in several
isolated mergers (NGC 3597: Lutz 1991; NGC 7252:
Whitmore et al 1993; NGC 4038/4039 (The Antennae): Whitmore 
\& Schweizer  1995),
such objects have not been found in cooling flow cD/gE galaxies, with
the possible exception of NGC 1275.
The situation for NGC 1275 is quite controversial, since there is both a
strong cooling flow and evidence for a recent merger in the form of
shell-like features in the optical and IR.
Holtzman et al (1992) argued against the blue
clusters having formed from the  cooling flow because there is no apparent
correlation between colour and magnitude.  Richer et al (1993) found a
large scatter in the colours of the blue clusters based on ground-based
photometry, which they used to argue for a cooling flow origin.  Faber
(1993) has questioned the Richer et al photometry, pointing out that some
objects are simultaneously blue in B$-$V and red in V$-$I, and
vice versa.  Better photometry is needed to resolve this issue.  The
spectrum of the brightest blue object in NGC 1275 (Zepf et al 1995b) is
equally consistent with a merger or a cooling flow origin, given its
age ($\sim$ 1 Gyr) and apparent solar metallicity.

Deep multicolour photometry of globular clusters in more cooling flow
cD/gE galaxies will help a good deal.  Clusters formed from cooling
flow gas should have a unimodal colour/metallicity distribution 
(although one could imagine contrived `punctuated' models, where cooling flows
and associated star/cluster formation are interrupted from time to time
by subcluster mergers or other processes), whereas clusters formed in
one or more merger events should have multimodal distributions
(e.g. Ashman \& Zepf 1992).

\noindent{\bf 3:}~~{\it The metallicity of 
the X-ray gas is 3--5 times higher than the
[Fe/H] $\simeq$ $-$1 typical of globular clusters in gE/cD galaxies.}

\noindent 
At first glance, it seems that we are
actually comparing two different things: X-ray gas
metallicity at the {\it current epoch}, and
globular cluster metallicities 10--15 Gyr ago.  
However,
a consensus seems to be developing that the ICM iron originated in cluster
ellipticals, ejected through SN-driven winds (e.g. Renzini et al 1993).
With such models, several authors have concluded that the bulk of the
ICM enrichment occurs at early times ($\sim$ 10 Gyr) over a {\it very short
time} (a few 10$^8$ years; e.g. David, Forman \& Jones 1991; Renzini
et al. 1993; Elbaz, Arnaud, \& Vangioni-Flam 1995).  
Thus, if cooling flows have produced
significant numbers of globular clusters, the majority of clusters 
are expected to have metallicities comparable to (or higher than,
if cluster self-enrichment is important) the {\it present}
ICM metallicity,  since the gas enrichment occurs so rapidly.
Instead, the ICM gas typically has [Fe/H]=$-$0.5 to $-$0.3
(e.g. White et al 1994), while 
the mean metallicity of clusters in gE/cD galaxies is typically 
[Fe/H]=$-$1 to $-$0.5. 
Note that 
ICM enrichment models are very uncertain, with the relative role of 
type I and type II SN being quite controversial, and the past SN
rate being essentially unknown.  A key issue to investigate is the
expected {\it spread} in metallicity of ICM gas in SN models, and
to compare with larger datasets of globular cluster metallicities.
An interesting finding is that the gas iron abundance decreases
slightly with ICM temperature, and it will be interesting to see if
similar behavior is found for the globular clusters as well.

Secker et al (1995) have recently found that [Fe/H] $\simeq$ $-$0.3
for globular clusters in NGC 3311, quite comparable to the typical
ICM metallicity.  At the same time, ASCA data have shown that 
[Fe/H] $\simeq$ $-$0.8 for gas in NGC 1404 and NGC 4374
(Loewenstein et al. 1994).  In NGC 4636, [Fe/H]$_{gas}$ $\simeq$
$-$0.5 near the galaxy centre, but declines
to [Fe/H] $\simeq$ $-$0.9 at R $\sim$ 9$^{\prime}$.  If such 
metallicity gradients are common, then previous X-ray data
(which lacked the ASCA spatial resolution) have resulted in 
{\it over-estimates} of the ICM metallicity. 

\noindent{\bf 4:~~} {\it It is 
difficult to form objects with globular
cluster masses from cooling flows.}  

\noindent It is well established that most of
the gas condensing out of cooling flows cannot be forming stars with a
`normal' (i.e. Galactic) mass function, else the colours and spectra 
of cooling flow cDs and their halos would show this clearly
(e.g. Fabian 1994). 
Instead, if star formation is occurring, it is presumably biased
towards M $\leq$ 1 M$_{\odot}$.  This can be explained qualitatively
by the high pressures expected in cooling flows, which would 
produce a low Jeans mass at T $\leq$ 10$^4$ K (M$_J$ $\propto$
P$^{-1/2}$).  However, if magnetic fields are present, clouds can cool
below 
10$^4$ K with lower densities and pressures than they would have
in the absence of the field, and with a correspondingly higher
Jeans mass.  Unfortunately, it requires magnetic field strengths of
$\sim$ 500--1000 $\mu$G to produce 
M$_J$ $\sim$ 10$^6$ M$_{\odot}$ 
(Fabian, private comm),  and typical cluster magnetic fields 
as measured from Faraday rotation are only 1--100 $\mu$G
(e.g. M87: ~Owen, Eilek, \& Keel 1990;
A1795: ~Ge \& Owen 1993; A2029 and A4059:~Taylor, Barton, \& Ge 1994.);
however, the field strengths could be much higher on small scales.

In any case, the Jeans-mass thermal instability approach to globular
cluster formation is difficult to reconcile with current observations.
Most notably, (a) the distribution of clusters by {\it mass} shows no peak
mass which could readily be identified with a `universal' Jeans mass;
instead, it shows a simple power-law form $ N(M) dM \sim M^{-\gamma}$
(e.g. Harris \& Pudritz 1994; note that the GCLF in its conventional
form as number of clusters per unit {\it magnitude} shows a distinct
peak at $M_V \sim -7.5$, but this is an artifact of binning in log-luminosity);
and (b) the cluster mass distribution is similar in all galaxies over a huge
range in metallicity, i.e. the heavy-element composition of the protocluster
gas.  Cooling flows (dilute, hot inflowing gas) represent a much different
gaseous environment than would a typical early protogalaxy (with lots of
material in massive, cool clumps which are colliding and merging, perhaps
embedded in an ambient hotter medium).  Thus if cluster masses were determined 
by thermal instability we would expect cluster mass distributions in
cooling-flow galaxies to be much different than those in spirals or
normal ellipticals,
which is contrary to the observations.  A model more in accord with cluster
formation either at early or present-day epochs is that put forward by 
Harris \& Pudritz (1994),
whereby globular clusters form within supergiant molecular
clouds ($10^8 - 10^9 M_{\odot}$) that are supported by turbulence and weak
($10 - 100 \mu$G) magnetic fields rather than thermal pressure.

\noindent{\bf Summary}

On balance, the above discussion argues that the bulk of the globular
clusters in gE/cD were not formed from cooling flows, 
although 
some small fraction
of globular clusters may have formed in this way.  
In particular, it seems most unlikely that cooling flows are responsible
for the high--S$_N$ phenomenon.
While 
cooling flows are not forming significant numbers of globular clusters
at the present epoch, 
they may have played an indirect role 
at earlier times.
Perhaps, as discussed by Merritt (1984,1985)
and others, cD galaxies were formed in small groups/subclusters, and there
was time for cooling flows to develop.  Somewhat later, the rich clusters
we see today were formed from the mergers of these subclusters.  This 
merging process may have also triggered star and globular cluster formation
from the collected cooling flow gas.  

\section{Conclusions}

We have obtained photometry to a limiting magnitude of R $\simeq$ 25
for globular clusters in NGC 6166.  Our main conclusions are:

\bigskip

$\bullet$~~The {\it local} globular cluster specific frequency is
S$_N$ = 9, with a possible range between 5 and 18, which is
inconsistent with the earlier result of 
S$_N$ $\leq$ 4
found by Pritchet \& Harris (1990) from their B-band images, yet
is consistent with a local S$_N$ in a similar region of M87.

$\bullet$~~The mean B$-$R colour for 37 globular clusters is
1.26 $\pm$ 0.11, which corresponds to a mean [Fe/H] = $-$1 $\pm$ 0.4
using a calibration based on Galactic globular clusters.  This
mean cluster metallicity is inconsistent with the 
[Fe/H] $\sim$ $-$0.4 found for the X-ray gas by White et al. (1994).
Our result confirms and extends recent findings of significant 
scatter in any possible relation between
mean cluster metallicity and parent galaxy luminosity at the 
high-luminosity end.

$\bullet$~~There is no apparent trend between R mag and B$-$R, and
no apparent globular cluster colour gradient. 

$\bullet$~~The available data and theoretical/numerical models 
allow us to rule out cooling flows as a mechanism for forming 
significant numbers of globular
clusters in gE/cD galaxies, and cooling flows appear {\it not} to be
responsible for the high S$_N$ phenomenon.
However, some small fraction of globular
clusters may form in cooling flows, and more and better data are needed of 
globular cluster systems in cooling flow and other
environments.  There also remain several interesting
theoretical issues to explore (the metallicity evolution of ICM
X-ray gas and the expected spread in metallicity, and the role
of SN; numerical modelling of star/cluster formation in cooling
flows, particularly the role of magnetic fields; a better understanding
of cluster formation in merging galaxies). 

\section*{acknowledgements}
We would like to thank Alastair Edge, Andy Fabian, Keith Ashman,
and Jeff Secker for
valuable discussions.  Thanks also to Steve Zepf for sending his 
Monte-Carlo code, to Reynier Peletier for sending his surface
photometry, and to Jeff Secker for sending the data for 
Figure 7.  We appreciate the detailed comments of Steve Zepf, Keith
Ashman, and Jeff Secker  on earlier drafts of this paper.  We would
also like to thank the anonymous referee for suggestions which
led to an improved discussion of consistency checks in Section 2. 
The research of WEH and CJP is supported in part by the Natural
Sciences and Engineering Research Council of Canada.

\newpage

\end{document}